\documentclass[sigconf]{acmart}

\acmConference[MSR 2022]{MSR '22: Proceedings of the 19th International Conference on Mining Software Repositories}{May 23–24, 2022}{Pittsburgh, PA, USA
}

\usepackage{cleveref} 
\usepackage{xspace}
\usepackage{balance}
\usepackage{booktabs}

\usepackage{array}
\newcolumntype{C}[1]{>{\centering\arraybackslash}p{#1}}
\newcolumntype{L}[1]{>{\raggedright\arraybackslash}p{#1}}
\newcolumntype{R}[1]{>{\raggedleft\arraybackslash}p{#1}}

\newcommand{\osas}[1]{open source automated software}

\definecolor{airforceblue}{rgb}{0.36, 0.54, 0.66}
\definecolor{amethyst}{rgb}{0.6, 0.4, 0.8}

\newcommand{\Qone}[0]{
What characterizes automotive software projects in open source?}

\newcommand{\vehiclesoft}{in-vehicle software\xspace}
\newcommand{\Vehiclesoft}{In-vehicle software\xspace}
\newcommand{\tools}{tools\xspace}
\newcommand{\Tools}{Tools\xspace}

\begin{document}
\title{Painting the Landscape of Automotive Software in GitHub}

\author{Sangeeth Kochanthara}
\email{s.kochanthara@tue.nl}
\affiliation {
\institution{Eindhoven University of Technology}
\country{The Netherlands}
}
\author{Yanja Dajsuren}
\email{y.dajsuren@tue.nl}
\affiliation {
\institution{Eindhoven University of Technology}
\country{The Netherlands}
}
\author{Loek Cleophas}
\email{l.g.w.a.cleophas@tue.nl}
\affiliation {
\institution{Eindhoven University of Technology}
\country{The Netherlands}
}
\author{Mark van den Brand}
\email{m.g.j.v.d.brand@tue.nl}
\affiliation {
\institution{Eindhoven University of Technology}
\country{The Netherlands}
}

\renewcommand{\shortauthors}{Kochanthara et al.}

\begin{abstract}
The automotive industry has transitioned from being an electro-mecha\-nical to a software-intensive industry.
A current high-end production vehicle contains 100 million+ lines of code surpassing  modern airplanes, the Large Hadron Collider, the Android OS, and Facebook's front-end software, in code size by a huge margin.
Today, software companies worldwide, including Apple, Google, Huawei, Baidu, and Sony are reportedly working to bring their vehicles to the road.
This paper ventures into the automotive software landscape in open source, providing a first glimpse into this multi-disciplinary industry with a long history of closed source development. 
We  paint the landscape of automotive software on GitHub by describing its characteristics and development styles.

The landscape is defined by 15,000+ users contributing to  $\approx$600 actively-developed automotive software projects created in a span of 12 years from 2010 until 2021.
These projects range from vehicle dynamics-related software; firmware and drivers for sensors like LiDAR and camera;   algorithms for perception and motion control;  to complete operating systems integrating the above.
Developments in the field are spearheaded by industry and academia alike, with one in three actively developed automotive software repositories owned by an organization.
We observe shifts along multiple dimensions, including preferred language from MATLAB to Python and prevalence of perception and decision-related software over traditional automotive software.
This study witnesses open source automotive software boom in its infancy with  many implications  for future research and practice. 
\end{abstract}

\maketitle

\section{Introduction}
Today,  automotive is a software-intensive industry~\cite{staron2017automotive, ebert2017automotive}. 
The latest innovations in this 5 trillion dollar industry\footnote{https://www.carsguide.com.au/car-advice/how-many-cars-are-there-in-the-world-70629} (including automated driving, intuitive infotainment, and electrification) depend less on mechanical ingenuity and more on software innovations.
In 2020, the software in a car and hardware it runs on is estimated to cost from \$4,800 up to \$10,650.\footnote{https://www.eetimes.com/projections-for-rising-auto-software-cost-for-carmakers/} By 2030, this cost is expected to double forming an estimated 50\% of the total car cost.\footnote{https://www.statista.com/statistics/277931/automotive-electronics-cost-as-a-share-of-total-car-cost-worldwide/}

The recent entry of the automotive industry in Open Source Software (OSS) 
is a land-marking change for an industry primarily driven commercially and dependent heavily on protecting their intellectual property.
This exposes the automotive software industry, consisting of original equipment manufacturers (or car makers in short),  their different tiers of suppliers, and tool vendors to a wide network of contributors worldwide, in addition to interesting and relevant projects.
To understand what exists and what opportunities this landmark change can offer, this study explores the \emph{landscape of automotive software projects in OSS}, as seen on GitHub.

Many studies have explored the landscape of OSS, albeit for different domains.
There are studies on AI-ML software~\cite{gonzalez2020state}, software from large tech companies~\cite{han2021empirical}, and even specific application domains like video games~\cite{murphy2014cowboys} and bots~\cite{wessel2018power}. 
To this, we add automotive software 
with its distinctive and unique blend of non-safety critical, safety critical, and infotainment software, bunched together into a single system. 
We investigate:

\vspace{1em}
\begin{center}
 \emph{\Qone}   
\end{center}

\vspace{1em}

We explore the following two dimensions: 
\\
\emph{(1) Categories \& characteristics:} We identify what types of automotive software projects are open sourced and compare them to each other.
 We also compare the automotive projects to non-automotive projects. 
Further, we explore the characteristics of automotive projects (e.g., size and maturity of the field) and their stakeholders (e.g., key players and affiliations).
\\
\emph{(2) Software development styles:} We investigate different aspects of software development like collaboration (e.g. types of contributors, their contributions and interactions) and   contribution style (e.g. independent vs. dependent).

Our analyses are based on $\approx$600 automotive and a similar count of non-automotive projects on GitHub created in a span of 12 years from 2010 to 2021. Our main contributions are: 
\begin{itemize}
\item A manually curated, first of its kind dataset  of actively developed automotive software and their classification along four popular dimensions including safety-critical software and tools~\cite{anonymous}. 
This dataset facilitates the replication of this study and  future  explorations into automotive software. 
\item A characterization of automotive software including its temporal trends, 
popularity, programming languages, user distributions, and  development activities.
\end{itemize}
To the best of our knowledge, this study is the first in presenting the automotive software landscape in open source. Insights presented in this study are 
 relevant for the field growing at a fast pace and yet little is known from a software engineering perspective.

The rest of the paper is organized as follows: Section~\ref{sec:method} presents our design choices for data collection and analysis. 
Section~\ref{sec:results_1} and~\ref{sec:results_2}
present our findings and insights along with our approach to derive these insights. 
Section~\ref{sec:implications} presents the implication of this study for automotive and software engineering research and practice. 
We review the threats to validity in Section~\ref{sec:threats}, describe the related research in Section~\ref{sec:related_work}, and conclude the paper in Section~\ref{sec:conclusions}.

\section{Study Design} 
\label{sec:method}

Our choice of GitHub for the exploration of the automotive software landscape  is motivated by the sheer volume of open source software projects hosted on the platform, and its prevalence worldwide.
In 2021 alone, 64 million new repositories were created, with more than 73 million contributors from over 200 countries around the globe and 84\% of the Fortune 100 companies using GitHub.\footnote{https://octoverse.github.com/}

There are three parts to this investigation.
First, we define automotive software and propose criteria to distinguish automotive software from general software systems; and criteria to identify general repositories serving as the baseline for comparison. 
The metadata of the two sets of selected repositories are used for the second and third part.
In the second part, we present descriptive statistics of the repositories (in \Cref{sec:results_1}) while in the third part we explore user statistics as well as contribution patterns (in \Cref{sec:results_2}).
For the second and third parts, we derive insights from the automotive domain and compare it against the baseline.
Particularly, we mine archival data via the GitHub API (using PyGithub - a python wrapper for GitHub API search\footnote{https://pygithub.readthedocs.io/en/latest/introduction.html}) for the second part. 
We further enrich this data with the GHTorrent data~\cite{Gousi13} 
for the third part.
Generally, our study design takes inspiration from recent landscape studies relating to OSS (e.g.,~\cite{gonzalez2020state,han2021empirical}).

\subsection{What is Automotive Software?}
There are many definitions of automotive software prevalent in different scientific communities (e.g., ~\cite{haghighatkhah2017automotive,
ebert2017automotive,
dajsuren2019automotive}).
Some common elements of these definitions are: 
(a) the software that forms part of a vehicle, 
(b) the software that interacts with a vehicle via APIs or other similar mechanisms, and (c) the software specifically used for creating (a) and (b)~\cite{ebert2017automotive, broy2007engineering, goswami2012time,haghighatkhah2017automotive}.
A more detailed characterization of automotive software is presented in \Cref{sec:results_1}.

\subsection{Identify Automotive Software Projects}
To identify a specific type of software projects on GitHub, conventional methods like topic modelling~\cite{hindle2011automated, panichella2013effectively} are found to be inefficient~\cite{gonzalez2020state}.
Another approach uses the `topics' feature on GitHub.\footnote{https://github.blog/2017-01-31-introducing-topics/}
Topics are labels defined by a project or suggested to a project (by GitHub) that can be used to discover a network of similar repositories.\footnote{https://github.com/topics}
Our preliminary manual analysis 
showed that unlike  
previous study~\cite{gonzalez2020state}, 
several automotive repositories did not use GitHub's `topics' feature.
Therefore, in addition to looking at `topics' to identify repositories, we searched GitHub for specific keywords which if found in the `README' file are likely to identify an automotive software repository.

To identify automotive software using the `topics' feature of GitHub, first we defined seed terms.
We choose `automotive', `automobile', `drive', `driving', `vehicle', `vehicular', and `car' as the seed terms.
To capture a range of related terms, we transformed the seeds terms to their base terms. For example, `automo' for automobile and automotive.
Likewise, the other keywords became:  driv, vehic, and car.
Using these base terms, we composed a search string excluding the terms that are not related to automotive software.
Examples are google-drive, e-commerce, and device-driver related topics.
Our final (4) search queries were:

\begin{quote}
\emph{
automo,\\ vehic,\\ driv NOT driven NOT drives NOT license \\NOT google-drive  NOT linux-driver,\\ car NOT cart   NOT card NOT caro \\NOT carp NOT care}
\end{quote}

Using these search queries we identified topics which collectively defined the search space for automotive software repositories.
In total, we identified 2,797 topic labels. 
We manually analyzed each topic to decide whether it is related to automotive software or not.
If a topic label was not informative, we looked at the name and description of the top 10 repositories linked to the topic to make the decision.
Ultimately, we identified 286 topics and selected all their linked repositories.
A complete list of the topics (along with its search term) is available as a part of our replication package~\cite{anonymous}.

Further, to identify relevant repositories that do not use topics, we selected the top five topic 
results based on repository count (from the  286 topics in the prior step),  from each of the four search queries, that are  selected in the prior step (yielding a total of 20).
For a better signal-to-noise ratio in the search results, we removed the most common terms (e.g., car, cars) which resulted in 12 terms. 
We searched for these terms in the `README` file of repositories which do not use `topic' labels, in order to identify additional repositories.
Notably, 
only up to 50\% of the repositories relating to automotive were found using the `topics' feature and 301 out of 584 selected automotive repositories did not use this feature. 
Note that in each of the above manual analysis steps, a random sample and borderline cases were analyzed by two researchers independently, to ensure rigor and repeatability.

\subsection{Selection and Elimination Criteria}
\label{sec:criteria}
To curate a representative sample of active projects, we apply the following filtering criteria (inspired by~\cite{gonzalez2020state}): 
\begin{enumerate}
\item[\emph{Size:}] The size of a repository should be greater than 0 KB.
\item[\emph{Popular:}] Stars and forks are indicators of the popularity of a repository. To collect a representative sample of repositories (and not just the popular ones), we select repositories with at least 5 forks OR 5 stars.
\item[\emph{Activity:}] We use commits as a proxy of development activities and select repositories where the last commit was in 2021, a criterion for selecting actively developed projects.
\item[\emph{
Data:
}] The repository data should be available via the GitHub API. 
The above four criteria when applied to the shortlisted software repositories, resulted in a subset of 1981 repositories.
\item[\emph{Content:}] To gauge whether a repository is an automotive software one or not, the first author manually examined the project title, description, and README file 
based on the following inclusion and exclusion criteria. \\
\emph{Inclusion criteria}
\begin{itemize}
    \item 
    Select automotive-specific software
    \item 
    Select software that aids in the development of auto\-motive-specific software 
    \item 
    Select software related to on-road vehicles only 
    \item  
    The text is written in English and has a README file
\end{itemize}
\emph{Exclusion criteria}
\begin{itemize}
    \item 
    Repositories that are not automotive related or relate to (automotive) sales and marketing, tutorials, course projects, bachelor and master theses,  documentations, data-sets, toy cars, games, traffic infrastructure, maps, and ones that do not directly interact with vehicles. 
\end{itemize}
\end{enumerate}
We adopted a conservative approach for selecting repositories.
This means that cases which fall in a grey area were excluded.
For the repeatability of the procedure, another researcher with experience in conducting empirical software engineering independently classified a subset of randomly selected repositories (approximately 100) using the above inclusion and exclusion criteria.
The inter-rater agreement between the two classifications was 0.83 as calculated using Cohen's Kappa~\cite{kvaalseth1989note} indicating an almost perfect agreement.
The two researchers discussed their disagreements until a decision was reached. 
In the end, we identified 585 automotive software repositories.

\subsection{Identify Baseline Repositories}
To compare our insight against a baseline, we needed actively developed repositories  
that are not automotive-related.
Our first choice was reusing the baseline from 
a related prior study~\cite{gonzalez2020state}.
This dataset, however, had three issues: 
(1) does not contain recent repositories (created after mid-2019), 
(2) systematically excludes AI-ML repositories, and
(3) represents most popular repositories which are not necessarily representative of general software projects.
To mitigate these concerns, we created our baseline with the following characteristics.
First, we identified actively developed projects using the same criteria (size, popularity, activity, and data availability) as for the automotive software projects (refer to Section 2.3).
The only deviation we made is selecting repositories with five or more stars and forks.
This decision was made to mitigate the practical implementation limits of the search API. 
Then, for each year (from 2010 until 2021), we sub-sampled repositories proportional to the percentage distribution of all the  actively developed GitHub repositories over the years, and selected based on most recent activity from each sub-sample.
We selected repositories such that their aggregate count is closer to 600 repositories.
To avoid overlap with the automotive software, we excluded repositories with the terms \emph{automotive, car,} and \emph{vehicle}.
Our resulting dataset had 566 repositories as baseline.

\subsection{Data Analysis}
There are two parts to our data analysis:
(1) We report descriptive statistics on the selection of automotive and baseline repositories. 
We describe the types of automotive software systems and how they relate to baseline software systems.
This part of our analyses is 
based on the meta-data extracted using PyGithub.
For details, refer to Section~\ref{sec:results_1}.
(2) We offer deeper insights into development styles by combining insights from PyGithub and GHTorrent~\cite{Gousi13}.
Since GHTorrent dataset contained developement data only upto July 2021 (we collected data using PyGitHub in December 2021), some of the repositories from  PyGitHub based data was not available in GHTorrent. 
Consequently,  we were left with 
436
out of 585 automotive repositories and 503 out of 565 the baseline repositories. 
Refer to Section~\ref{sec:results_2} for deeper implementation details along with obtained insights.
 \section{Categories \& Characteristics}
 \label{sec:results_1}

 This section presents the types of automotive software available on GitHub and their characteristics.
 First, we introduce the different ways to classify automotive software. 
 The next subsection presents our findings and the distinctive characteristics of automotive software with reference to the comparison set of general software systems. 
This analysis 
is based on  the 584 automotive repositories (extracted using PyGitHub) created in a span of 12 years between 2010 and 2021.
The most recent one was created on 30th December 2021.

   \subsection{Approach}
 Informally, automotive software can be defined as: 
 (1) the software that runs or interacts with a vehicle; and (2) the tools to support different life cycle stages (e.g., development, validation \& verification) of the software that runs or interacts with a vehicle.
 We refer to the above two categories as \emph{\vehiclesoft} and \emph{\tools}, respectively.

\vspace{1em}
\par \emph{\Vehiclesoft:
} 
In literature, there are many ways to categorize \vehiclesoft. 
We use the following two schemes:\\
(1) \emph{Safety critical \& safety critical based on application:}  \emph{Safety critical software} is defined as the software that carries out tasks, which if not properly performed, could lead to human injury, death, or harm to the environment~\cite{nasa,hanssen2018safescrum,mcdermid2013software,knight2002safety,kochanthara2016revert,kochanthara2016revert2}.  
During the manual classification of automotive software, we noticed that in addition to safety critical and non-safety critical software systems, there is a third type of software systems: \emph{safety critical based on application}.
These software systems can be safety critical depending on the (intended) application context.
For instance, a software system for perception  is safety critical when used in fully automated driving (i.e., without an active human driver).
In this case, any failure, malfunction, or unintended function of the perception software system can lead to a crash, injury to the traffic participants, and harm to its surroundings.
The same system when used as a driver-warning system, in which human driver is in charge, can be classified as non-safety critical.
In this scenario, the responsibility of maneuvering the vehicle is with the human driver. 
We classified such software repositories as \emph{safety critical based on application}. \\
(2) \emph{Broy's classification:} In 2007, Broy et al.~\cite{broy2007engineering} classified automotive software  into the following 5 categories: 
(a) \emph{Human Machine Interface (HMI), multimedia, and telematics} related software; 
(b) \emph{Body/comfort software}, for instance, the software for controlling various aspects of car doors; 
(c) \emph{Software for safety electronics}, that are  hard real-time, discrete event-based software with strict safety requirements; 
(d) \emph{Powertrain and chassis control software}, which include control algorithms and software for controlling the engine; and 
(e) \emph{Infra\-structure software}, like software for diagnosis and software updates. 

Since 2007, the field of automotive software and software systems in general has evolved. 
For example, highly accurate image recognition with (relatively) lower computational power was demonstrated using neural networks in 2012~\cite{krizhevsky2012imagenet}. 
Automotive software has advanced in perception systems and automated decision making which makes fully automated driving possible.
While other aspects of automated driving like drive-by-wire are captured in the current classification, this aspect of perception and decision making, however, is not. 
We extend Broy's classification by adding a sixth category: (f) \emph{perception and decision software}. 
Perception and decision software includes any software that contributes to the perception (understanding the surroundings of a vehicle) and decision making (e.g., deciding actuation, steering, and brake), for any level of automated driving (i.e., driver assistance, partially automated, and fully automated). 

\vspace{1em}
\par \emph{\Tools:} 
Industry standards (e.g., ISO26262~\cite{ISO26262,kreiner2017integrated}) define multiple stages, such as validation \& verification, in the automotive  life-cycle. 
We consider all software repositories that offer tools for one or more stages of an automotive life-cycle, in this category. 
For brevity, we exclude from this classification, the stages (and hence the corresponding tool) for which we had exactly one repository. 
The selected repositories fall into the following four categories:
(1) tools for development, (2) tools related to simulation or emulation, (3) tools for validation \& verification, and (4) tools for diagnostics. 

Please note that traditionally tools relating to simulation (and emulation) are considered part of validation \& verification. However, with the advent of neural networks, many simulation tools are used in training (developing) neural networks. Therefore, we study these tools separately here.

To ensure a rigorous and repeatable classification of automotive repositories into the above-mentioned categories, the first author and another researcher with experience in empirical software engineering, independently classified a subset of randomly selected repositories and borderline cases.

In the subsequent subsection, we report the distribution of automotive software based on the above classification to offer an overview of the types of automotive software open sourced on GitHub. 
We continue venturing into these popular and prominent classes of automotive software throughout the paper.

To characterize automotive software, we report descriptive statistics on automotive software repositories and compare them against the comparison set of non-automotive repositories,  also from GitHub.
Our analyses highlights four key areas, starting with the \emph{temporal trends and evolution of the repositories} on GitHub. This analysis indicates the maturity and growth of the field. 
Next, we discuss the \emph{ownership of the automotive software} (users versus organization) indicating the key players of the field and how they are shaping the landscape of automotive software.
Along the same lines, we continue exploring \emph{popular automotive software} in terms of development activities (inferred from fork count) and in general (using stars and subscribers).
We conclude with an exploration into the choice of \emph{primary programming languages} used by different automotive software, as identified using the above categories.

\subsection{Findings}

\noindent
\textbf{
Genesis
 - The beginning \& temporal trends:} 
In 2010, the first still actively developed automotive project, \emph{Veins}, was created in GitHub (26th-April-2010).
This vehicular network simulation framework defined the entry of automotive software development into GitHub, 
 marking a turning point for an industry traditionally closed source in the past 50 years of its software use.

\begin{figure}
    \centering
    \includegraphics[width=0.4\textwidth]{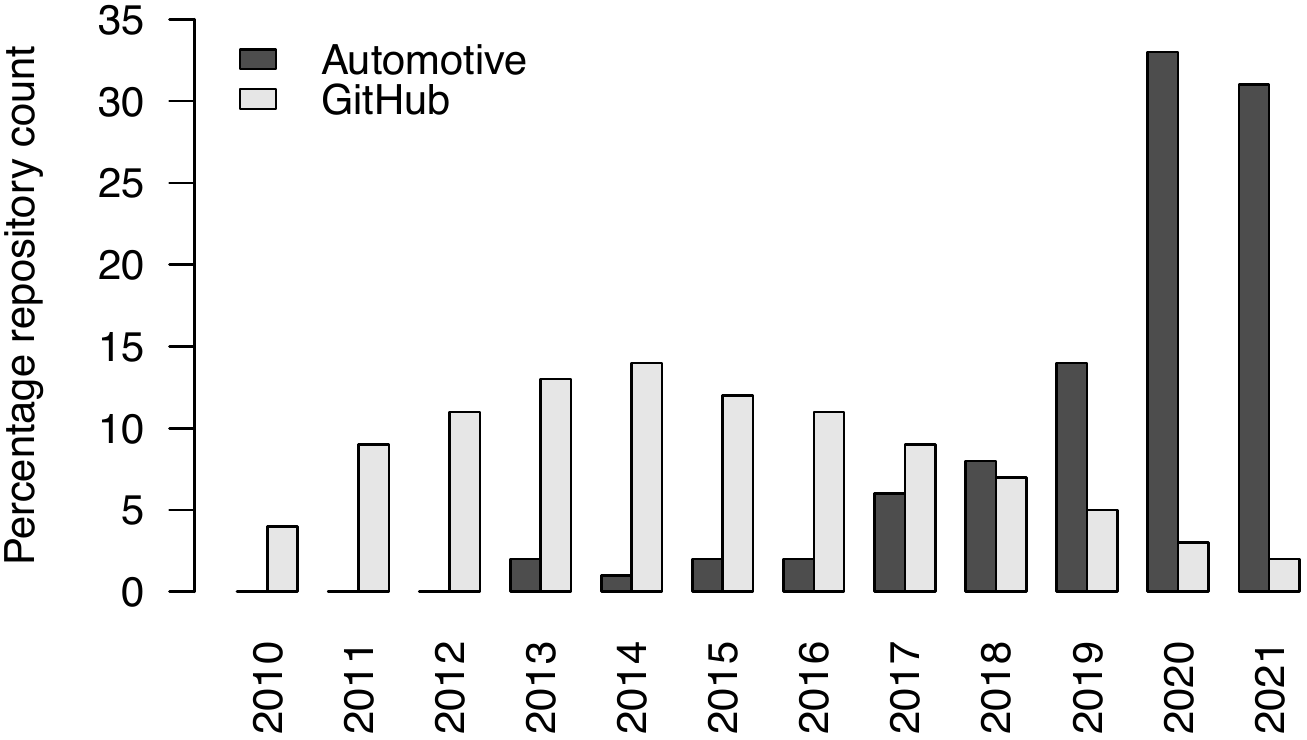}
    \caption{Temporal evolution of actively developed automotive software repositories with reference to all actively developed repositories in GitHub, between 2010 and 2021.
    }
    \label{fig:temporal_trend}
\end{figure}

Since then more repositories are added each year to a total of 584 (actively developed) automotive software repositories in a span of 12 years. 
 \Cref{fig:temporal_trend} presents the percentage distribution of the automotive software repositories (based on their creation year)  with reference to the total actively developed repositories in GitHub.
The temporal trend suggests that from 2018 to 2019 the percentage growth of automotive software has doubled which again doubled from 2019 to 2020. 
In comparison to the actively developed repositories across GitHub (which peaked in 2014), automotive software is still in its infancy and expected to grow in the future.
\vspace{1em}

 \setlength{\fboxrule}{0.7pt}
 \noindent\fbox{%
    \parbox{0.465\textwidth}{%
        \textbf{Origin \& temporal trends:}  
        \emph{Veins} - a simulation tool, is the first automotive software repository created by a user in 2010 that is still actively developed. 
        This study witnesses open source automotive software boom in its infancy.
    }%
}
\vspace{1em}

 \noindent
\textbf{The dawn of new opportunities - Ownership:} 
In 2010, no organization was  developing any automotive software in GitHub that is still actively developed.
In the next 6 years (2011-2016), small organizations, enthusiast groups, and non-profits organizations ventured into open source; owning 14 software projects. The first organization owned (still actively developed) project is \emph{Open-Vehicle-Monitoring-System}, an \vehiclesoft from the enthusiast group Open Vehicles. 

The year 2017 marked the entry of big players to open source. 
This year Baidu, the Chinese search engine company, created project \emph{Apollo} - a full software stack for fully automated driving.
The other big players, namely, Amazon, Intel, Microsoft, and Udacity joined soon after, each of which open sourced one of their tools to build automated driving solutions. In our data set, one in three (or 194 out of 584) automotive software repositories is owned by an organization. 
Cumulatively, we are looking at 163 organizations and 343 users owning at least one repository.

 The development of automotive OSS today is spearheaded by tool vendors, academic, and industrial research labs with more than 30 repositories owned by academic research groups. 

The only car maker in automotive software on GitHub is Toyota with two repositories, a tool and an \vehiclesoft.
Please note that we might have missed the different tiers of suppliers to car makers since there is no straightforward way to identify suppliers from the GitHub meta-data.

Currently, the top 5 organizations working on automotive software (in terms of repository count) are: VITA lab at EPFL (5 repositories), LG Silicon Valley Lab (4), MathWorks Open Source and Community Projects (4), AutonomouStuff (3), and CARLA (3).

\vspace{1em}
 \setlength{\fboxrule}{0.7pt}
 \noindent\fbox{%
    \parbox{0.465\textwidth}{%
        \textbf{Repository ownership:} One in three automotive software repositories is owned by an organization. The field witnesses high participation from academic and industrial (tool vendors) research labs  with only one car maker (Toyota) at the forefront.
    }%
}
\vspace{1em}

 \noindent
 \textbf{In the catbird seat - 
 Popularity:} 
  The top 5 popular automotive projects from organizations  based on the three indicators of popularity (stars, forks, and subscribers; see \Cref{tab:top_5}) are simulation tools (4 in count), followed by in-vehicle software on perception and decision systems (fizyr/keras-retinanet) and the automated driving stack Apollo.
 
The popular user repositories on automotive software are more diverse, including both  \vehiclesoft and \tools. 
The \vehiclesoft in the top 5 relate to (a) HMI and telematics and (b) perception and decision. The \tools in the top 5  relates to (a) the development of perception and decision-related software and (b) diagnostics. 

\begin{figure}
    \begin{center}
    \includegraphics[width=0.4\textwidth]{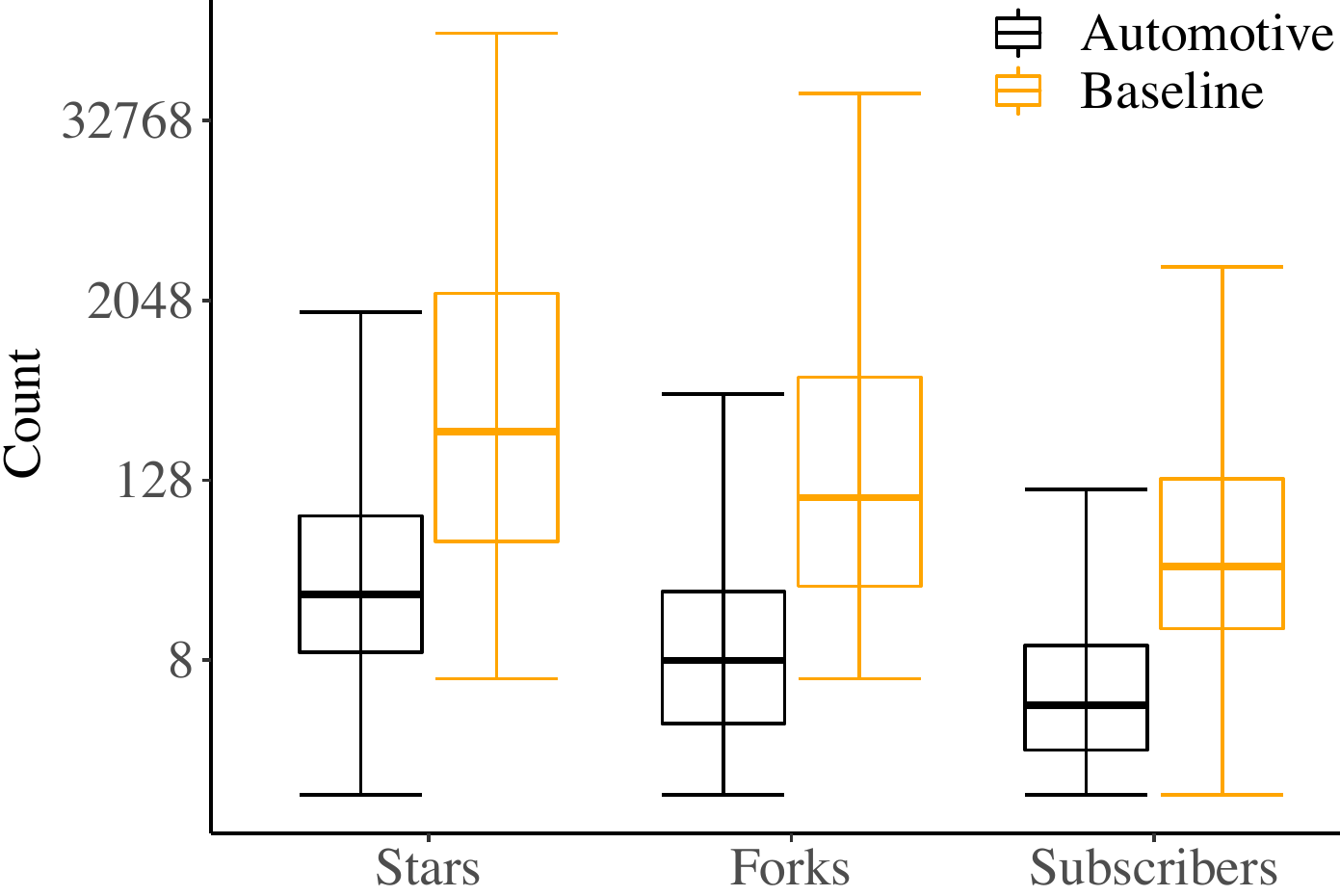}
        \end{center}
    \caption{Distribution of the popularity of automotive repositories (in terms of stars, forks, and subscribers) with reference to the comparison set (outliers removed)}
    \label{fig:popularity}
\end{figure}

 \Cref{fig:popularity} presents the distribution of stars, forks, and subscribers across automotive software along with the baseline repositories. 
 Generally, projects have more stars, than forks and subscribers. 
This is the same for automotive and baseline repositories.
Notably, automotive software is far less popular than the baseline software systems, further reinforcing the notion of infancy of the field.
The differences in the distribution among the automotive and baseline repositories are statistically significant as calculated using the Mann-Whitney-Wilcoxon Test~\cite{bergmann2000different}, a non-parametric test for two independent data samples, calculated at p-value$<$0.05. 
The median numbers of stars, forks and subscribers for automotive repositories are 24, 9, and 4 respectively, while the median for baseline repositories are 297, 121, and 36, respectively.
Here, we would like to remind the readers that despite our attempts at selecting a representative sample of projects as baseline, our dataset might be somewhat biased towards more popular repositories (see threats to validity  in Section~\ref{sec:threats} for details), skewing the distribution further.
Even within automotive repositories, organization-owned repositories is at least twice as popular as user-owned software projects. 
  
\vspace{1em} 
  \setlength{\fboxrule}{0.7pt}
 \noindent\fbox{%
    \parbox{0.465\textwidth}{%
        \textbf{Popularity:} 
        Automotive software as a field is less popular than general software on GitHub. 
        Apollo, Baidu's automated driving software stack,  is currently the most popular automotive repository. 
        Generally, organization-owned software projects are twice as popular as user-owned projects.
        }%
}
\vspace{1em}

\begin{table}[!htp]\centering
\caption{Top 5 popular organization and user repositories (based on subscribers, forks, and stars) and their count}\label{tab:top_5}
\scriptsize
\begin{tabular}{lrrr}\toprule
&Organization  & User \\\midrule
&\multicolumn{2}{c}{\textbf{Top 5 repositories based on subscriber count}} \\\hline
1 &ApolloAuto/apollo (1103) &stanleyhuangyc/ArduinoOBD (175) \\
2 & microsoft/AirSim (597) &timdorr/tesla-api(119) \\
3 &carla-simulator/carla (236) &Smorodov/Multitarget-tracker(110) \\
4 &udacity/self-driving-car-sim (231) &cedricp/ddt4all(81) \\
5 &autoas/as (145) &fr3ts0n/AndrOBD(68) \\ \\
&\multicolumn{2}{c}{\textbf{Top 5 repositories based on fork count}} \\ \hline
1 &ApolloAuto/apollo(8013) &MaybeShewill-CV/lanenet-lane-detection(772) \\
2 & microsoft/AirSim(3557) &Smorodov/Multitarget-tracker(569) \\
3 &carla-simulator/carla(2128) &stanleyhuangyc/ArduinoOBD(486) \\
4 &fizyr/keras-retinanet(1964) &timdorr/tesla-api(474) \\
5 &udacity/self-driving-car-sim(1414) &karlkurzer/path\_planner(355) \\ \\
&\multicolumn{2}{c}{\textbf{Top 5 repositories based on star count}} \\\hline
1 &ApolloAuto/apollo(19954) &MaybeShewill-CV/lanenet-lane-detection(1707) \\
2 & microsoft/AirSim(12590) &Smorodov/Multitarget-tracker(1636) \\
3 &carla-simulator/carla(7100) &timdorr/tesla-api(1549) \\
4 &fizyr/keras-retinanet(4252) &poodarchu/Det3D(1220) \\
5 &udacity/self-driving-car-sim(3595) &yangyanli/PointCNN(1200) \\
\bottomrule
\end{tabular}
\end{table}
  
\noindent
 \textbf{Genera - Types of automotive software:} 
Broadly, there is an abundance of \vehiclesoft (375) in comparison to tools (233).
A detailed distribution of the types of automotive software (both \vehiclesoft and \tools) is presented in \Cref{tab:category}. 
Note that a repository can belong to multiple categories. Therefore, the sum of the repository count in all the categories can be greater than the actual repository count. 
 More details on the classification of individual repositories is available in the replication package~\cite{anonymous}. 

Within \vehiclesoft, most repositories relate to perception and decision related software. 
Notably, Broy's classification~\cite{broy2007engineering} for \vehiclesoft is a small part (108 out of 375) of the whole. 
Of these 108, HMI (71) and infrastructure (38) are the top categories, and are primarily developed by users.

The development of safety-critical software in open source is still in its initial stages (21 repositories). 
Although, many software repositories belong to safety-critical based on application category (approximately 60\% of \vehiclesoft).
 Most of the safety-critical software relates to perception and decision-based software intended for use in fully automated driving systems. 
These include neural-network based semantic segmentation, path planning, and object, pedestrian, and intent detection related software.

The category `tools' in automotive software is dominated by simulators and related software. 
This is evident in the top five automotive software from industry (see Table~\ref{tab:top_5}), three of which are simulators.
For tools overall, there is near to equal ownership from users (100) and organizations (133). 
We see a similar trend in the ownership of \emph{development tools} (32 from organization and 31 from users) and \emph{validation \& verification tools} (18 from organizations and 22 from users). 
The only exception is \emph{diagnostic tools} which are five times more prominent among users (28) than organizations (5). 
Other than the above, we also notice a small number of automotive software repositories (32) relating to (driver) safety (like drowsiness detection) and security (tools for security testing or \vehiclesoft for the security of the vehicle).

\vspace{1em}
 \setlength{\fboxrule}{0.7pt}
 \noindent\fbox{%
    \parbox{0.465\textwidth}{%
        \textbf{Types of automotive software:} The most popular type of automotive software developed open-source is \emph{\vehiclesoft} (375 repositories) followed by \emph{\tools} (233 repositories). 
        Within \vehiclesoft, perception and decision software are most popular while in \tools, simulations are prominent.
        Traditional vehicle software and safety critical software are underrepresented in open source.
    }%
}
\vspace{1em}

\begin{table}[!htp]\centering
\caption{Types of automotive software on GitHub and their distribution. }\label{tab:category}
\small
\begin{tabular}{
L{0.25\textwidth} 
C{0.045\textwidth}
C{0.045\textwidth} 
C{0.045\textwidth}
}\toprule
Category& Org & User &Total 
\\\midrule
\textbf{\Vehiclesoft} & \textbf{97} & \textbf{278}  &\textbf{375} \\ \\
\hspace{0.2cm} Safety-critical &11 &10 &21 \\ \\
\hspace{0.2cm} Safety-critical based on application &57 &167 &224 \\ \\
\hspace{0.2cm} \emph{Extended Broy's classification} 
& & & \\
\hspace{0.4cm} HMI, multimedia, \& telematics &17 &54 &71 \\
\hspace{0.4cm} Body/comfort software &10 &8 &18 \\
\hspace{0.4cm} Software for safety electronics &8 &4 &12 \\
\hspace{0.4cm} Power train and chassis control software &10 &7 &17 \\
\hspace{0.4cm} Infrastructure software &9 &29 &38 \\
\hspace{0.4cm} All Broy's categories combined &23 &85 &108 \\ \\
\hspace{0.4cm}perception and decision software &68 &180 &248 \\\\ 
\textbf{\Tools} &\textbf{100} &\textbf{133} &\textbf{233} \\ \\
\hspace{0.2cm} for development &32 &31 &63 \\
\hspace{0.2cm} for validation \& verification &18 &22 &40 \\
\hspace{0.2cm} Simulation (and emulation) related &48 &52 &100 \\
\hspace{0.2cm} for diagnostics &5 &28 &33 \\
\bottomrule
\end{tabular}
\end{table}

 \noindent
\textbf{Two worlds; two languages - Languages:} 
Automotive software is developed in 33 primary languages and 96 languages when considering all the languages for development. 
The most popular programming language is Python with 291 projects using it as a primary language and up to 415 projects using it as one of the languages.
The top 5 primary programming languages 
are Python (291), C++ (98), C (33), Jupyter Notebook (33), and MATLAB (30). 
Technically Jupyter Notebook is not a programming languages, rather a blend of text and code. We do not make any assumptions in the programming languages used inside the Notebooks rather considered them according to the tagging by GitHub. 
Organizations generally prefer (based on repository count) Python (82), C++ (37), C (16), MATLAB (10), and Jupyter Notebook (7) for their projects.
Notably, users also prefer the same languages but in slightly different order: Python (209), C++ (61), Jupyter Notebook (26), MATLAB (20), and C (17).
Most safety critical software is written in C++ (14), followed by Python (4), MATLAB (1),
and C (1).

The distribution of programming languages across projects shows a shift from MATLAB as preferred language of development~\cite{altinger2014testing} to Python. 
Likewise, traditionally most safety critical software were developed in C or Ada
which has now shifted to C++ in GitHub~\cite{kastner2020safety}. 

\vspace{1em}
 \setlength{\fboxrule}{0.7pt}
 \noindent\fbox{%
    \parbox{0.465\textwidth}{%
        \textbf{Languages:} The preferred language for open source automotive software development has shifted to Python (291 repositories) from MATLAB (30 repositories). Similarly, safety critical software development has moved from C or Ada to C++. 
    }%
}
\vspace{1em}
\section{Software Development Style}
\label{sec:results_2}
Building on the insights derived in the previous section,
this section delves into the user distribution, types of development activities, and the choice of development models in automotive software.
We compare it against the baseline to understand the unique characteristics of automotive software, if any. 
Note that since this analysis combines data acquired using PyGithub (in December 2021) with GHTorrent's data (data available until July 2021), we missed repositories which do not exist on GHTorrent. Further, depending on the development activities of individual repositories and missing data in GHTorrent dataset, the total count of the repositories may vary across different analyses.

 \subsection{Approach}

 \emph{User distribution:} 
 In this section, we explore the types and distribution of users across projects. 
 We study two types of users: external and internal~\cite{gonzalez2020state}, based on their activities in automotive software.
 \emph{Internal users} contribute directly to the development of a project by making changes to the actual software (commits) and moderating the decision to include/exclude the proposed changes (like merging and closing pull requests and closing issues).
 \emph{External users}, on the other hand, contribute indirectly by requesting features, reporting issues, and commenting.
 We believe that investigating the distribution of internal and external users across projects indicates how a community works. 
 For deeper insights, we also explore changes in contribution patterns, if any, across organization and user projects.

A natural next step to understand developer contribution and collaboration patterns is to examine developer roles (e.g., maintainer, or reporter) and their distribution.
However, given the small community size 
and limited development activities, it is infeasible to offer meaningful insights and conclusive statistical analyses. Therefore, we do not report collaboration patterns.

\vspace{1em}
\par
 \emph{Development activities:}
Development activities on GitHub can be broadly classified into  commits,  issues, and  pull requests. 
Issue events indicate participation from the broader user base (beyond contributors) requesting additional features or indicating problems. 
Participation in issue events indicate how the users of the software interact with developers, influencing its development.
The next group of development activities are pull requests which indicate a relatively stronger influence on the software by proposing changes for inclusion into the software system or its associated artifacts.
These activities log the decisions to include or exclude proposed changes. 
Finally, a commit is an even more involved activity dealing with the technical aspects of creating desired changes in the software.\footnote{https://docs.github.com/en/get-started/quickstart/github-flow}
Here too, we explore whether project ownership influences the development activity patterns across projects.\\

\par \emph{Development models \& autonomy:} Finally, we analyze the choice of development model in automotive software.
There are two types of development models in GitHub: (1) shared repository model and (2) fork \& pull model.\footnote{https://docs.github.com/en/pull-requests/collaborating-with-pull-requests/getting-started/about-collaborative-development-models}
The two models are different in the level of autonomy of contributors. 
In the shared repository model, an author can merge their proposed code changes themselves, indicating their autonomy.
The fork \& pull model implies that the changes proposed by an author are reviewed by a maintainer. 
In this model, an author is dependent on the actions of a reviewer for the decision to include or exclude the proposed changes. 
To study the level of autonomy or dependence in a project, we aggregate the distribution of pull requests and commits for which the author merged the changes (self-merge) versus other contributor (other-merge). 
We refer to the projects with more self-merges as practicing a shared-repository model and fork and pull model otherwise.

 \subsection{Findings}
\begin{figure}
    \centering
    \includegraphics[width=0.4\textwidth]{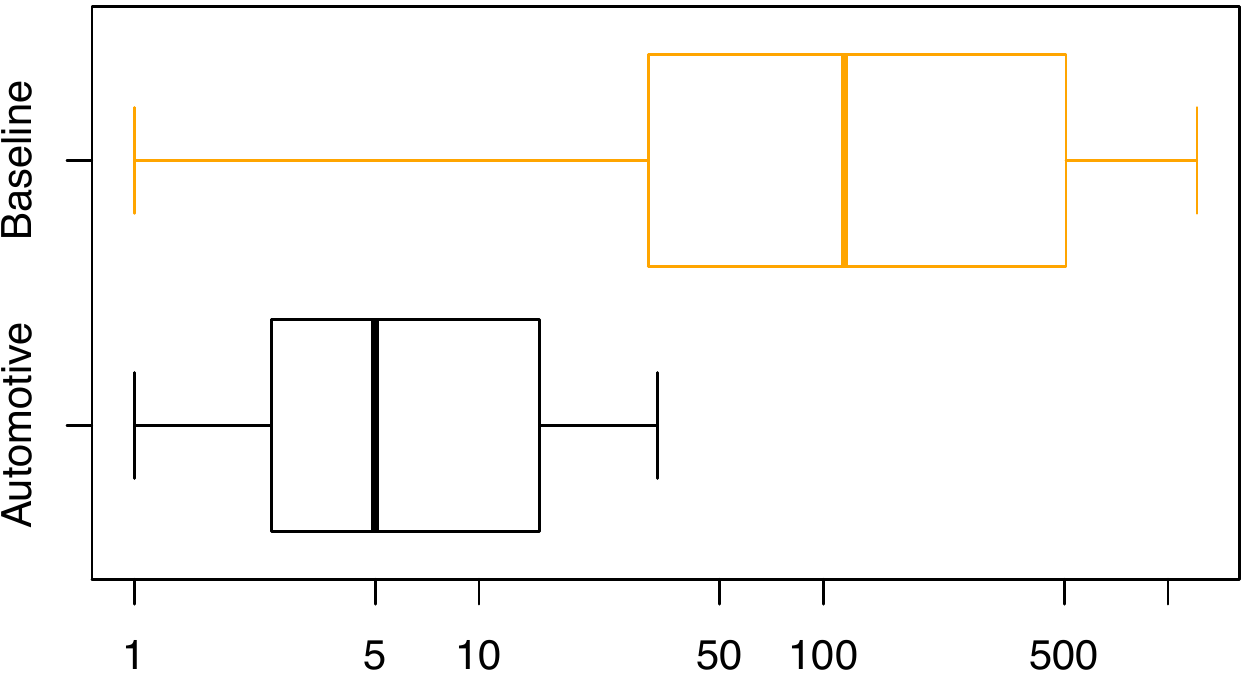}
    \caption{User distributions across repositories compared with baseline (outliers omitted)}
    \label{fig:user_distribution_all}
\end{figure}

 \textbf{The real stars - Users \& their distribution:} 
 All the automotive repositories cumulatively have 15,260 unique users where as the baseline set of repositories have 439,032 unique users.
 In automotive software, the median count of users per repository is 5 while 115 for general projects from our baseline (refer to~\Cref{fig:user_distribution_all} for distribution).
 The two distributions are significantly different as calculated using the Mann-Whitney-Wilcoxon test at p-value$<$0.05.
  Here again, we warn our readers that our baseline is somewhat skewed towards actively developed and popular software systems. 
  Therefore, the differences may appear larger than they are. 
  Note that our identification of individual contributors relies on unique user identifiers from GitHub. However, one individual can have several unique identities~\cite{robles2005developer}. Consequently, the actual number of users might be lower than the reported count.

\begin{figure}
    \centering
    \includegraphics[width=0.37
    \textwidth]{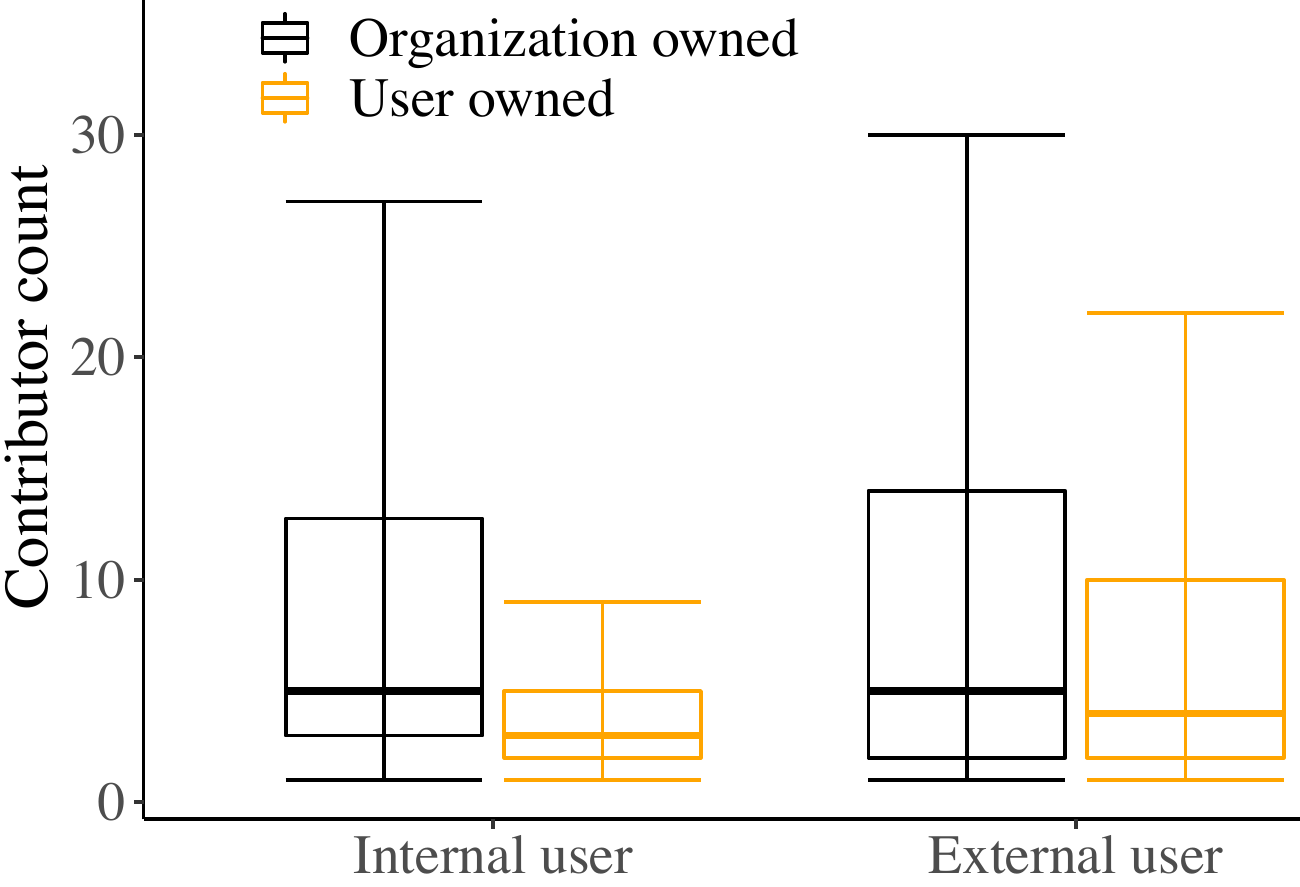}
    \caption{Internal and external users across automotive repositories owned by users and organizations (outliers omitted) }
    \label{fig:user_distribution_owner}
\end{figure}

Looking at the distribution of users in automotive software, each project has a median count of 3 internal and 5 external users. 
When we further segregated the user distribution on the ownership type (see \Cref{fig:user_distribution_owner}), we observe that organizations have more users per repository. The two distributions are different as calculated using the Mann-Whitney-Wilcoxon test at p-value$<$0.05.
Organizations recorded a median of  6 internal  and 8 external users in comparison to 3 and 5 respectively for the user owned automotive repositories.

\vspace{1em}
 \setlength{\fboxrule}{0.7pt}
 \noindent\fbox{%
    \parbox{0.465\textwidth}{%
        \textbf{User distribution across repositories:} Open source automotive software has a small developer community with a median of 5 users per repository. 
        Notably organization repositories solicit more participation internally and externally in comparison to the user repositories.
        }%
}
\vspace{1em}

 \noindent
 \textbf{Abiogenesis - Development activities:} We notice that most common development activities (based on the median count of activities) in automotive software are in the form of commits (32), followed by issues (9), and then pull requests (6).
 We further investigated the distribution of development activities based on the ownership and the types of automotive software. 
 
\begin{figure}
    \centering
    \includegraphics[width=0.4\textwidth]{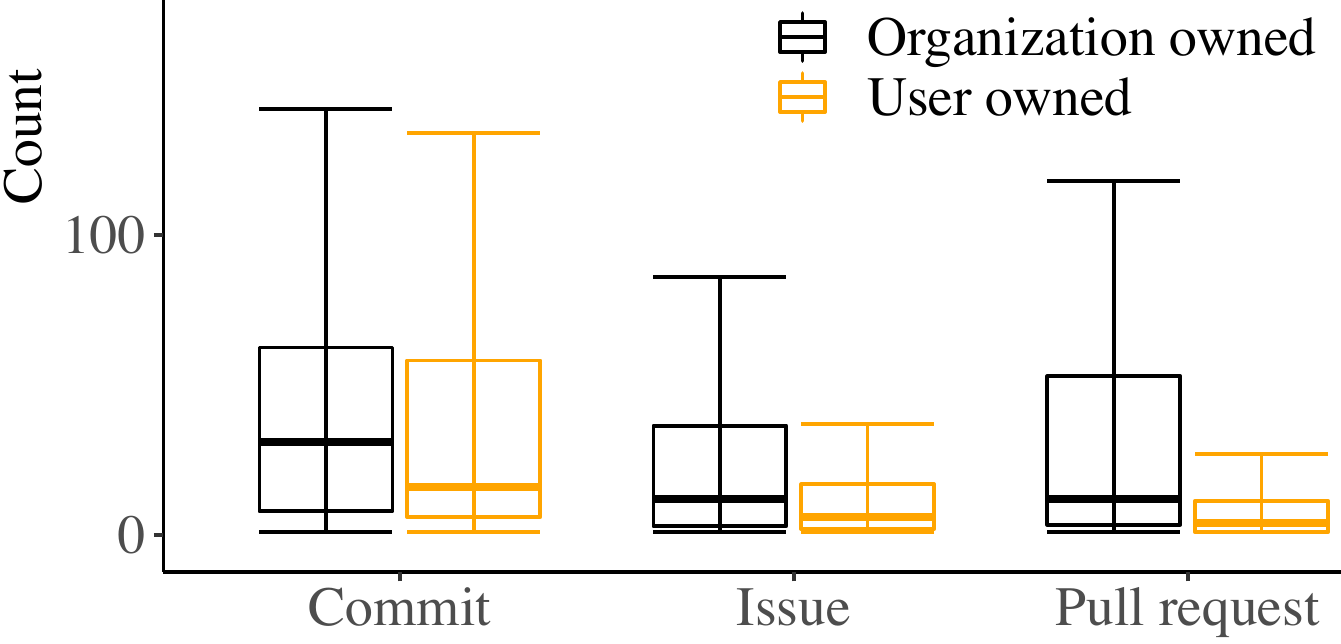}
    \caption{Development activities in organization owned automotive repositories versus user owned repositories (outliers omitted)}
    \label{fig:icp_ownership}
\end{figure}

 \Cref{fig:icp_ownership} presents the development activities of automotive software repositories owned by organizations  in comparison with the repositories owned by users. 
 Generally, organization projects have more development activities (in terms of median) than user projects do, although there are a few user projects where the commit activity level matches that of the organizational repositories.
 This indicates variability among user projects with extremes in the distribution of development activities. 
 Please note that the two distributions are different as calculated using the Mann-Whitney-Wilcoxon test (p-value$<$0.05).

\begin{figure}
    \centering
    \includegraphics[width=0.38\textwidth]{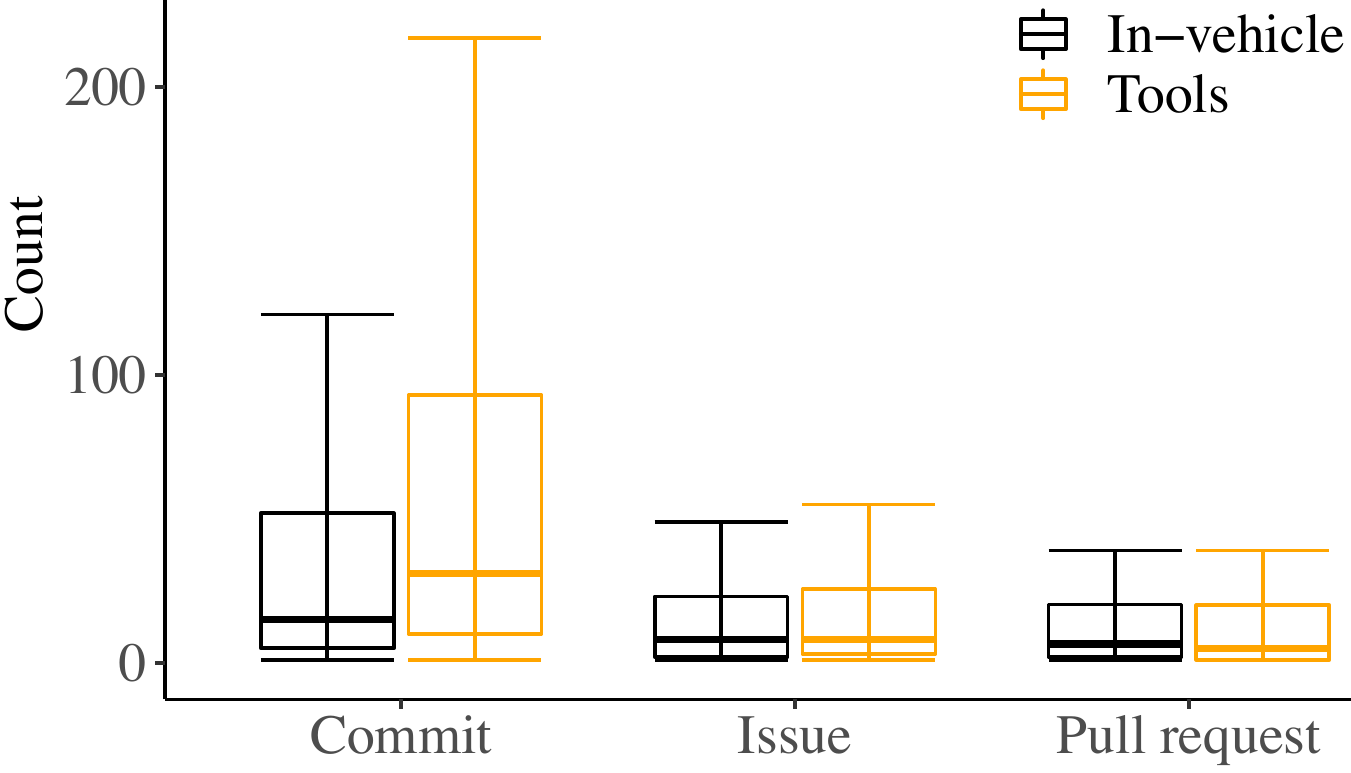}
    \caption{Development activities for \vehiclesoft versus \tools (outliers omitted)}
    \label{fig:icp_broad_category}
\end{figure}

 A comparison of the development activities across \vehiclesoft and \tools is shown in \Cref{fig:icp_broad_category}. 
 Here, the development activities across issues (p-value = 0.29) and pull requests (p-value = 0.40) are comparable with major differences in the commits (p-value = 0.0013).
 These p-values are calculated using the Mann-Whitney-Wilcoxon test such that a p-value$<$0.05 indicates differences in the distribution and no differences otherwise.
 We notice that there are more commit activities in \tools in comparison to \vehiclesoft.
 We believe that this difference is attributed to the higher participation of tool vendor organizations in automotive software.
 To remind, there are 100 out of 233 organization owned repositories in the \tools category versus 97 out of 375 in the \vehiclesoft category.

\begin{figure}
    \centering
        \includegraphics[width=0.4\textwidth]{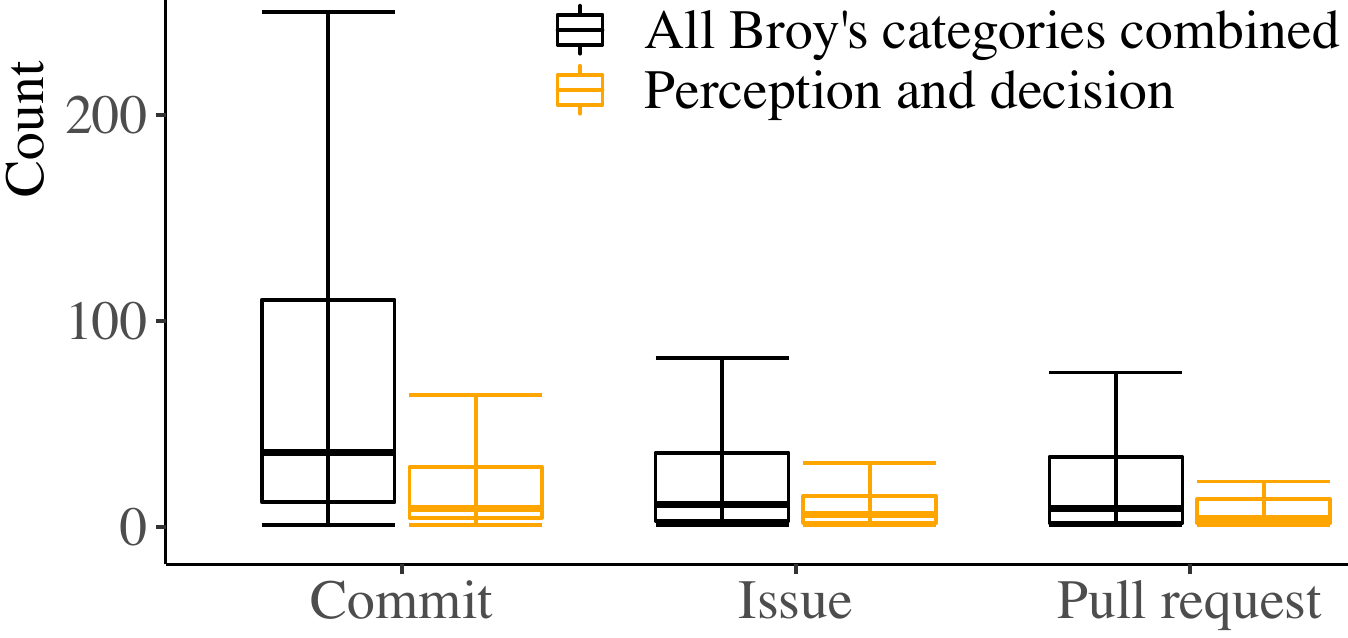}
    \caption{Development activities for perception and decision related \vehiclesoft versus other \vehiclesoft (outliers omitted)}
    \label{fig:icp_perception}
\end{figure}

 We also explored the distribution of development activities in perception and decision related software, which form a majority of the \vehiclesoft (with sizable number of repositories for statistical comparison) with respect to the development activities with the rest of repositories in \vehiclesoft. 
\Cref{fig:icp_perception} shows that traditional software is more actively developed in terms of commits, issues, and pull requests than perception and decision related software.
The differences in the distribution of development activities on repositories belonging to all the Broy's categories combined versus repositories belonging to  perception and decision software, are significantly different as measured using the Mann-Whitney-Wilcoxon test at p-value$<$0.05.
This means that while perception and decision related software are more in count,
they have fewer development activities than traditional software.

\vspace{1em}
 \setlength{\fboxrule}{0.7pt}
 \noindent\fbox{%
    \parbox{0.465\textwidth}{%
        \textbf{Development activities:} The organization-owned repositories attract more contribution across all categories. However, there are smaller number of user owned repositories matching the level of development  activities as in organization owned repositories. Tools are more actively developed than   \vehiclesoft. Within  \vehiclesoft, traditional \vehiclesoft is more actively developed than perception related software.
        }%
}
\vspace{1em}

\noindent
\textbf{Guardians - Development model:} We classified each project as a shared repository or fork-and-pull development model by looking at their collaboration patterns. 
We noticed that while 41 repositories followed the shared repository model, 182 repositories practice the fork-and-pull model. 
This indicates that fewer projects are autonomously developed and a majority 
of the development teams (contributors of a repository) do not have autonomy. 
This observation matches the baseline with fork-and-pull being the most prominent model and also prior work on other software sub-communities~\cite{gonzalez2020state}.

\vspace{1em}
 \setlength{\fboxrule}{0.7pt}
 \noindent\fbox{%
    \parbox{0.465\textwidth}{%
        \textbf{Development model:} Automotive software---despite being a small development community---mostly follows the fork-and-pull model for its development activities. 
        }%
}
\vspace{1em}

\section{Implications}
\label{sec:implications}
While automotive open source software is still in its infancy (in comparison to our baseline), the field is 584 repositories and 15,260 contributors strong. 
With 
industry players entering the field and choosing to open source their projects, we believe that automotive software is the next promising area with an expected multi-fold growth in the next decade. 
In this section, we discuss how the insights from this study can be applied to research and practice.

\subsection{Research}
Academia is a prominent contributor of automotive software in GitHub with one of the top 5 organizations in terms of count of repositories owned.
We also identified 108 repositories linked to scientific articles in automotive software.
Given the close link of automotive software and academic research, in this section, we present the ways in which our study can inspire future research.

\emph{Manually curated automotive software.} To the best of our knowledge this paper is the first in presenting a manually curated dataset of automotive software in open source.
We have classified each repository using four categorization schemes namely (1) safety critical, (2) safety-critical based on application, (3) Broy's classification extended with perception and decision related software, and (4) \tools.
One immediate future research direction is strengthening the classification with inputs from experts in the automotive and related domains.
Also, our manually curated dataset along with the classification can be used (in training and testing) to automate the process of identifying and classifying automotive software by automated (algorithmic) approaches. Further, we 
believe that future research can use this classification for the characterization of automotive software 
and in-depth explorations into prominent software systems. 

\emph{Motivation to open source.} Relating to the types of automotive software open-sourced by companies, we noticed that organizations mostly open source their tools and some experimental projects. 
It will be interesting to see what types of companies come to open source and their motivation (similar to a recent study exploring motivations of Chinese companies to open source~\cite{han2021empirical}). 
While our results show that one in three automotive software repositories is owned by an organization, it is possible that at least some of the GitHub projects in large organizations might have started as personal projects because there was not yet a company policy on how to open source projects. Given the entry of organizations to open source in the automotive domain is still in its infancy, it is possible that some individual projects might actually belong to organizations and the transfer of ownership to the organization has yet to be made. This phenomenon in itself and the subsequent changes in our results form another direction to explore for future research.

\emph{Safety-critical.} Most automotive software in open source relates to perception and decision and tools. 
These software systems can be safety-critical depending on the application. 
Given the limited attention to safety-critical software in open source, it remains a future work to see whether the current and future automotive software in these categories are developed and/or tested according to safety critical standards. Also, our study forms a guide and baseline for future studies on open source software in (other) safety critical domains.

\emph{Multi-disciplinary software versus general software.} Automotive software developed in GitHub is multi-disciplinary
in nature.
We observed repositories that model vehicle dynamics;\footnote{\url{https://github.com/TUMFTM/velocity_optimization}} 
develop firmware and drivers for sensors like LiDARs and Camera;  develop algorithms for perception and motion control; and repositories on complete operating systems integrating the above.
It will be interesting to see differences in the contribution and collaboration patterns of such multi-disciplinary software projects in comparison to general software systems. 

\subsection{Practice}
This study shows that the industry is more interested in open sourcing tools (43\% organization-owned) in comparison to in-vehicle software (26\%).  Three possible explanations for this observation are: (1) the revenue stream for the tools is dominated by car-makers or their suppliers who will pay for their support irrespective of open-sourcing; (2) tools might be less intellectual property intensive than in-vehicle software and thus easier to open source without losing the edge to competition; and (3) in-vehicle software typically runs on less-standardized hardware (application-specific embedded hardware for various in-vehicle functionality) than tools. 
These explanations however, cannot be derived from the data used in this study and need to be validated.

In addition, our list of automotive repositories and their categorization can aid future research into identifying the stakeholders in each category, their motivations to participate in open source, and other domains where these tools are used. Our study also provides the first list of automotive software tools and in-vehicle software in open source, with the former available for practitioners to use (and contribute to) and the latter 
to learn from.
Further, our data and insights can be used to identify (a) software for reuse, (b) attract talent and/or increase the adoption of software and standards, and (c)  new directions, companies, and trends in the automotive domain. 
Three potential implications of our findings for practice are discussed below.

\emph{Language of choice.} Automotive software in open source is now developed mostly in Python, replacing MATLAB as reported in prior studies (see Table 12 in~\cite{altinger2014testing}).
A similar trend of C++ domination replacing C or Ada for safety critical software development~\cite{kastner2020safety} is seen.
Our advice to the readers interested in venturing into automotive software is to consider these findings while choosing programming languages.

\emph{Companies.}
Prior study has shown that one reason for companies to open source their software is to attract talent and internationalization~\cite{han2021empirical}.
We believe that start-up automotive companies can benefit
by open sourcing their projects.  

\emph{Safety certification \& car makers.} Safety certification of  vehicles is obligatory to allow them on road. The current trend of more software dependent functionalities in vehicles is challenging for certification bodies. These bodies can benefit from open software stacks. Such a change can subsequently encourage car makers to contribute to open source software. Our study can be used by the certification bodies 
 to get insights on the characteristics of automotive software developed in open source. To car makers our study offers  trends in the open source automotive software.

\section{Threats to Validity}
\label{sec:threats}

\textit{Construct Validity:} 
There are threats to the representativeness of the automotive software systems selected for analyses.
To mitigate this concern, we identified automotive software repositories using two approaches (using topics and keyword search in README files) and adopted best practices for selecting actively developed software systems~\cite{gousios2017mining, kalliamvakou2016depth, munaiah2017curating}.
That being said, we might have systematically missed the repositories that do not use the search terms, uses different topic labels, misses README file, or a meaningful description. 
Also, in case of doubt, we discarded a repository, i.e.~we followed a conservative approach. 
The same threat applies to the categories of automotive software systems presented in Section~\ref{sec:results_1}.

 For counting unique contributors, we used their GitHub identifier and counted every user with a distinct identifier as a unique user. However,  prior studies have shown that one individual can have several identities~\cite{robles2005developer}, thus the actual number of users might be lower than the reported count.

Our baseline set of repositories might be skewed towards more popular repositories. For representativeness, we sub-sampled repositories based on the number of actively developed repositories created every year. 
However, given the smaller size of automotive repositories as compared to the total number of actively developed repositories in GitHub, the random selection from each sub-sample based on recent activity might have resulted in the selection of popular repositories as the baseline.

Another threat is the introduction of researcher bias in the manual selection of repositories and the classification of the automotive software.
While these threats cannot be eliminated, we tried to minimize them by (a) clearly documenting the inclusion-exclusion criteria for the selection of repositories, and (b) using an independent rater for a subset of the repositories. 
For the classification of software, we borrowed the definitions from literature for reference. 

Our insights into development styles rest on the GHTorrent dataset. 
The activities that are not present in the dataset are systematically excluded from our analyses~\cite{aranda2009secret}.

\textit{External Validity:} 
This paper is based on the publicly available software repositories on Github.
While GitHub is a popular and widely used platform, there are other platforms (e.g., Gerrit and Phabricator) with their distinctive characteristics, private projects hosted on GitHub, and closed source systems. They might add a different perspective to the automotive software landscape. 
We leave it to future research to pick these topics to improve the generalizability of the findings here.

\section{Related Work}
\label{sec:related_work}
Related studies around automotive software landscape in GitHub can be divided into two parts.
The first part focuses on the literature on automotive  software and its engineering. 
In the second part we explore studies offering characterization of other software communities.

\subsection{Automotive software}

There are many studies on automotive software, exploring the different dimensions of the topic. Some areas focused in the last five years (as identified using Google Scholar search) include automotive software architecture ~\cite{staron2017automotive, kugele2018data,kochanthara2021functional,kochanthara2020semi,kochanthara2021summary},  AI-based solutions ~\cite{falcini2017deep, salay2018using}, model-based solutions~\cite{schlatow2017towards, obergfell2019model, zoppelt2018sam} and blockchain~\cite{dorri2017blockchain}.
These studies touch on aspects such as complexity~\cite{kugele2017service}, safety~\cite{kugele2017service}, security~\cite{zoppelt2018sam, oka2018shift}, privacy~\cite{dorri2017blockchain}, and testing~\cite{oka2018shift, parthasarathy2020controlled} that are relevant for automotive software. 

Another line of research on automotive software is in terms of their development and development processes~\cite{
katumba2014agile, staron2017automotive, dajsuren2019automotive}. These studies focus on the different steps in automotive software development~\cite{staron2017automotive, dajsuren2019automotive} and applicability of process models like agile development~\cite{katumba2014agile} to automotive industry. 
That said, to the best of our knowledge, there is little to no study characterizing automotive software development and its process in terms of its development activities (like pull requests, issues, and commits) on closed or open source software systems. 

\subsection{
Non-Automotive Software Landscape}
Even though the  software development process or its characterization is not studied for automotive domain, characterization of other software engineering communities has been presented in literature.
The software engineering communities whose characteristics are explored in the past can be classified based on application domain~\cite{murphy2014cowboys,gonzalez2020state,wessel2018power} and those based on  other factors like geography and closed source~\cite{han2021empirical,han2021empirical}. 
We present four such studies that have explored software landscape from different perspectives and have inspired our study.  

The most recent exploration is on the open source software systems developed by large Chinese technology companies namely Baidu, Alibaba, and Tencent~\cite{han2021empirical}.
Unlike the open source software studied in general, this exploration is regional.
It presents a characterization of open source software developed by Chinese technology companies, their objectives for open sourcing, and a comparison to other software systems~\cite{han2021empirical}. 

The second study explores a decade of ML and AI software systems developed in open source~\cite{gonzalez2020state}. 
The study characterizes the trend of ML/AI evolution in addition to their collaboration and autonomy, and contrasts it against the general software systems~\cite{gonzalez2020state}. These two studies are our primary inspirations.

Another study characterizes video game development and how it is different from traditional software development~\cite{murphy2014cowboys}.
Based on interviews, the study identifies differences between the two types of software system and how researcher can help~\cite{murphy2014cowboys}.

Finally, studies on bots explore its use, and how these special software systems can help the development of other software systems~\cite{wessel2018power}.

Along these lines, this study offers an exploration into the landscape of open source automotive software projects.
Open sourcing is a recent phenomina in automotive industry as shown in the temporal trends in Section~\ref{sec:results_1}.  
Taking inspiration from the previous studies and combining elements from many sources, we quantitatively analyze the repository data.
We hope that like the previous studies, the findings from our study inspires future research and improve the state of automotive software development.

\section{Conclusions}
\label{sec:conclusions}
This study presents a landscape of automotive software projects publicly available on GitHub.
We identified and categorized $\approx$600 automotive repositories grounded in definitions from literature and well-defined empirical methods.
We also identified a similar number of non-automotive projects for comparison.
We analyzed the origin, temporal trends, key players, popularity of projects, languages for development, user distribution across repositories, and development activities.
We also present, a first of its kind, manually curated dataset of automotive projects and a comparison set of non-automotive projects, for replication and future research.

For an industry traditionally being in closed source in its half a century history of software use, open sourcing software projects marks a landmark change.
This study shows that automotive domain is undergoing  a shift in multiple dimensions including the prevalence of automated driving software development, change in preferred language from MATLAB to Python, and entry of software companies and start\-ups to the domain.

We foresee that the recent developments in software engineering, that enables automated driving, will further accelerate  open source automotive software development.
We believe that the software stacks for automated driving will benefit from perception and decision software  currently developed in open source.
Since these systems are developed independent of car makers, involving the open source community  
for the acceleration of their development, is a logical step.

\section{Acknowledgements}
\label{acknowledgements}

We thank Ayushi Rastogi for her role in the inter-researcher agreement, discussions on statistical tests, and help with the GHTorrent data.
Thanks are due to the MSR reviewers for their comments which resulted in a richer discussion section.
This work is a part of the i-CAVE research programme (14897 P14-18) funded by NWO (Netherlands Organisation for Scientific Research).

\balance
\bibliographystyle{ACM-Reference-Format}
\bibliography{ASL}

\end{document}